\documentclass[dvipdfm]{pasj00}
\usepackage{graphicx}
\draft
\newcommand{{\xmm}}{XMM-Newton}
\newcommand{{\chan}}{Chandra}
\newcommand{{\suz}}{Suzaku}
\newcommand{{\cxo}}{CXOU~J171405.7$-$381031}

\begin{document}
\SetRunningHead{\textsc{Sato} et al.}
{A new magnetar {\cxo}}


\title{Identification of CXOU~J171405.7$-$381031 as a New Magnetar with
{\xmm}}

\author{Takuro \textsc{Sato}\altaffilmark{1,2},
Aya \textsc{Bamba}\altaffilmark{3,1},
Ryoko \textsc{Nakamura}\altaffilmark{4},
and
Manabu {\textsc Ishida}\altaffilmark{1}
} %
\altaffiltext{1}{Institute of Space and Astronautical Science (JAXA),
3-1-1 Yoshinodai, Chuo-ku, Sagamihara, Kanagawa 252-5210, Japan}
\altaffiltext{2}{Department of Physics, Tokyo Metropolitan University, 1-1
Minami-Osawa, Hachioji, Tokyo 192-0397, Japan}
\altaffiltext{3}{Dublin Institute for Advanced Studies, 
School of Cosmic Physics, 31 Fitzwilliam Place, Dublin 2, Ireland}
\altaffiltext{4}{Earth Observation Research Center,
Japan Aerospace Exploration Agency,
1-1 Sengen 2chome, Tsukuba-shi, Ibaraki 305-8505, Japan}
\email{tsato@astro.isas.jaxa.jp, abamba@cp.dias.ie,
ishida@astro.isas.jaxa.jp}


%


\maketitle

\begin{abstract}

 We have observed the 3.8~s pulsar {\cxo} with {\xmm}, and discovered
 the significant $\dot P$ of $6.40\pm 0.14\times 10^{-11}$ s~s$^{-1}$
 from this source for the first time, with the aid of archival {\chan}
 data. The characteristic age (950~yr), the magnetic field strength
 (5$\times 10^{14}$~G), and the spin-down luminosity ($4.5\times
 10^{34}$ erg~s$^{-1}$) derived from $P$ and $\dot P$ lead us to
 conclude that {\cxo} should be identified as a new magnetar. The
 obtained characteristic age indicates that {\cxo} is youngest among all
 known anomalous X-ray pulsars,
which is consistent with the age estimation from the thermal X-rays
of the associated supernova remnant.
The ratio between 2--10~keV luminosity and spin-down luminosity
is almost unity,
which implies that
{\cxo} is the key source to connect
magnetars and traditional radio pulsars.

\end{abstract}

\section{Introduction}

Anomalous X-ray Pulsars (AXPs) have been distinguished from the other
type of pulsars by their peculiarity such as a long spin period of a
neutron star ($P =$2--12~s) and moderately bright X-ray emission with
$L_{\rm X} \approx 10^{33-35}$~erg~s$^{-1}$ that is in
 general much
greater than the spin-down luminosity of the neutron star 
($\dot{E} \approx 10^{31-34}$~erg~s$^{-1}$)\footnotemark[$*$].
\footnotetext[$*$]{http://www.physics.mcgill.
ca/$^\sim$pulsar/magnetar/main.html}
Since there
is no evidence of mass accretion, it has been thought that the X-ray
emission is replenished with magnetic energy of the neutron star. As a
matter of fact, the magnetic field strength, estimated on the basis of
the dipole radiation framework, of the AXPs with known $\dot P$ is all
in excess of the critical magnetic field $B {\rm c} = 4.4\times
10^{13}$~G, above which the differential energy of the neighboring
Landau levels exceeds the rest mass energy of an electron. The AXPs,
together with Soft Gamma-ray Repeaters (SGRs), are now classified as
so-called ``magnetars''
\citep{1992ApJ...392L...9D,1995MNRAS.275..255T,1996ApJ...473..322T}.
As of June 9, 2010, 10 AXPs and 9 SGRs have been
known\footnotemark[$*$].
It is, however, still unclear how such extraordinary nature are endowed
to the magnetars, compared with the other conventional pulsars. One of
the difficulties is that only a few magnetars are found associated with
host supernova remnants (SNRs) that provide us with information on their
progenitors, independent age estimation, and so on. 
Obviously we need
more magnetar samples with the SNR association.

{\cxo} is discovered in the course of identifying Galactic TeV sources.
\citet{2006ApJ...636..777A} discovered the TeV source HESS~J1713$-$381
with the atmospheric Cherenkov telescope H.E.S.S. and indicated its
association with the supernova remnant (SNR) CTB37B. Using {\chan} data,
\citet{2008A&A...486..829A} identified the point source {\cxo} in the
radio shell of CTB37B \citep{1991ApJ...374..212K}. Its location is
slightly offset ($\approx$\timeform{1'}) from the peak of the H.E.S.S.
brightness contour (see Fig.~2 of
\cite{2009PASJ...61S.197N}). \citet{2009PASJ...61S.197N} observed CTB37B
with {\suz} \citep{2007PASJ...59S...1M}, and found that its spectrum is
represented well by a power law with a photon index of $3.3\pm 0.2$. The
hydrogen column density to {\cxo} ($\approx$4$\times 10^{22}$~cm$^{-2}$)
is consistent with that of diffuse thermal emission of CTB37B, which
strengthens the association of {\cxo} to the SNR.
The distance to this SNR is estimated to be 10.2$\pm$3.5~kpc
\citep{1975A&A....45..239C},
and we cite this value in this paper.
These facts lead them to conclude that {\cxo} is probably a new AXP,
although limited time resolution of the XIS (8~s:
\cite{2007PASJ...59S..23K}) onboard {\suz} in the full-frame mode
and the ACIS (3.24~s: \cite{2000AJ....120.1426G}) onboard {\chan}
in the imaging mode
preclude them to detect pulsation. \citet{2010ApJ...710..941H} finally
discovered using the {\chan} cc mode observation data that {\cxo}
pulsates at the period of 3.82305$\pm$0.00002~s
on 2009 January 25,
which is well within the range of the AXP pulse
period. The pulse shape is sinusoidal with a pulse fraction of 31\%.
They also detected an excess emission above the power-law spectrum above
$\sim$6~keV, which is one of the common features among the AXPs.

To further strengthen the identification of {\cxo} as an AXP, it is
important to measure $\dot P$. The known $\dot P$ of AXPs exceeds
$10^{-12}$ s~s$^{-1}$\footnotemark[$*$], which is 
systematically larger than the other
rotation-powered pulsars. Furthermore, under the dipole radiation
assumption, we are able to estimate the strength of the magnetic field,
which is important to see if {\cxo} is a magnetar. In order to evaluate
$\dot P$, we have carried out an observation of CTB37B with {\xmm}
\citep{2001A&A...365L...1J}.
Monitoring X-ray flux is also important, since most magnetars show X-ray
time variability.
\citet{2010ApJ...710..941H} showed that
its flux changed between {\suz} and {\chan} observations,
although
{\suz} low spatial resolution prevented us to conclude
this source showed the time variability
since there could be contamination of diffuse nonthermal X-rays,
which is quite common in young SNRs \citep{2005ApJ...621..793B}.
%

In \S~2, we describe how the observation and data screening have been
carried out. In \S~3, we present the results of our timing and spectral 
analysis. 
We have clearly detected pulsation from the source. Compared with the
{\chan} pulse period \citep{2010ApJ...710..941H}, we detected $\dot P$
significantly. We calculate the characteristic age and the magnetic
field strength in \S~4 and argue that {\cxo} should be regarded as a new
magnetar.

\section{Observation and Data Reduction}

\begin{figure}[htb]
\begin{center}
\FigureFile(70mm,50mm){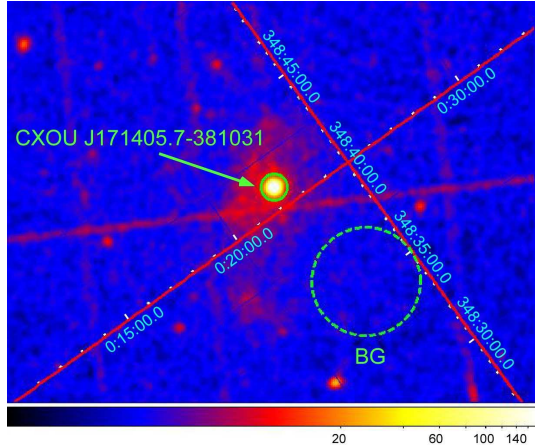}
\end{center}
 \caption{The {\xmm} pn image of CTB37B region in the 1--10~keV
 band. The image is smoothed with a Gaussian with $\sigma =$ 3
 pixels. The small green circle with a radius of \timeform{30''} is the
 region for extraction of the source photons, whereas the other dashed
 circle ($r = 2'$) is the region for background
 extraction.\label{fig:image}}
\end{figure}

The {\xmm} observation of the SNR CTB37B was carried out from 2010 March
17 13:16 (UT) to March 18 23:06 (UT). 
Our primary objective is timing analysis to determine the 
physical parameter of \cxo. Hence, we concentrate on
data taken with the EPIC pn \citep{2001A&A...365L..18S} whose time
resolution is 73.4~ms, which is much better than EPIC 
MOS (2.6 s; \cite{2001A&A...365L..27T}).
We have carried out the data analysis with
SAS ver 9.0.0. We have first checked the background flare using the
cleaned event file pipe-line-processed with the CCF dated on 2010 April
30 in the data package. We have extracted a light curve in the band
10--12~keV, and produced a GTI file that excludes time intervals with a
10--12 counting rate of $>$0.35 c~s$^{-1}$.
We then have revised the event file by applying this GTI file. 
As a result, the effective exposure time became 40.264 ks.
Using the new event 
file thus processed, we have constructed a pn image 
in the band 1--10~keV, as
shown in Fig.~\ref{fig:image}.
{\cxo} is clearly detected at $\ell$ = \timeform{348D50'51''.087}, $b$ =
\timeform{0D22'15''.820}, which is consistent with that from {\chan}
\citep{2008A&A...486..829A}.
The non-thermal diffuse emission extending to the south from {\cxo}
\citep{2009PASJ...61S.197N} is also detected.  The small green circle,
with a radius of \timeform{30''}, is the extraction region of the source
photons,
whereas the other dashed circle, with a radius of
\timeform{2'} is that for the background.
The intensity of
{\cxo} is 0.264$\pm$0.003 c~s$^{-1}$ in the 1--10~keV band after
subtracting the background.

%
\begin{figure}[htb]
\begin{center}
\FigureFile(80mm,60mm){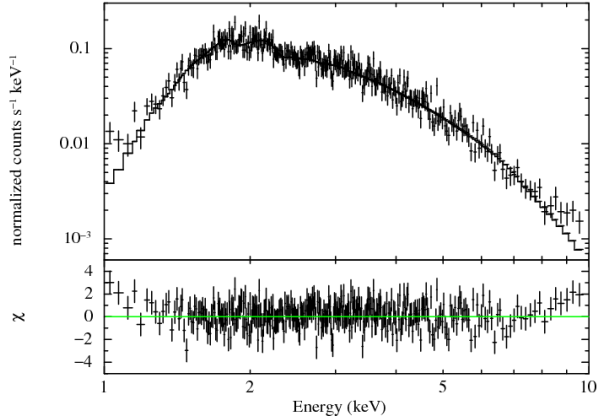}
\end{center}
 \caption{The {\xmm} pn spectra of {\cxo} and the two background
 regions. The background spectra are corrected for the aperture
 size.\label{fig:spectra}}
\end{figure}
%

Figure~2 shows the background-subtracted spectrum together with the
best-fit model and the fit residuals.
We can see deeply absorbed and hard emission.
We have fitted an absorbed power-law model to the data.
The metal composition of \citet{1989GeCoA..53..197A} is adopted
as the solar abundance for the absorbing material.
The fit is accepted with  $\chi^2$/d.o.f. of 372.34/341.
The best-fit photon index, the hydrogen column density,
and the observed and intrinsic flux in the 2.0-10.0~keV band are
$3.45^{+0.09}_{-0.08}$, $3.95^{+0.15}_{-0.14}\times 10^{22}$~cm$^{-2}$,
$(1.51\pm0.03)\times 10^{-12}$~erg~cm$^{-2}$s$^{-1}$,
and $(2.68\pm0.09)\times 10^{-12}$~erg~cm$^{-2}$s$^{-1}$, respectively
(the errors represent single-parameter 90\% confidence limit).
We can see positive residuals above $\sim$5~keV,
which is probably the hard tail
which is common among the magnetars \citep{
2007MNRAS.378L..44M,2008PASJ...60..237N,2009PASJ...61..109N,enoto2010}.
Detailed spectral analysis will be presented in the forthcoming
paper.


\section{Timing Analysis}

\begin{figure}[htb]
\begin{center}
\FigureFile(80mm,50mm){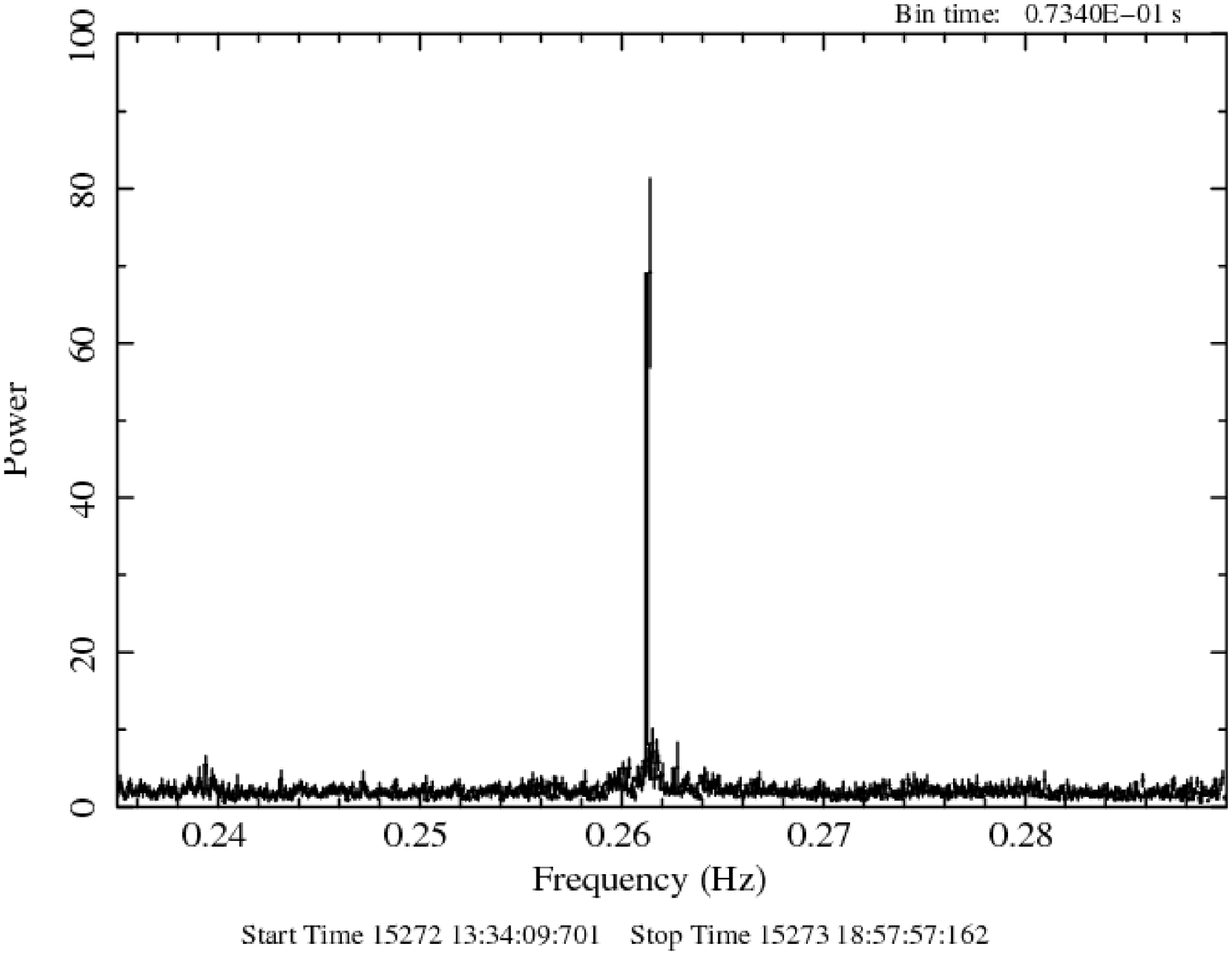}
\FigureFile(80mm,50mm){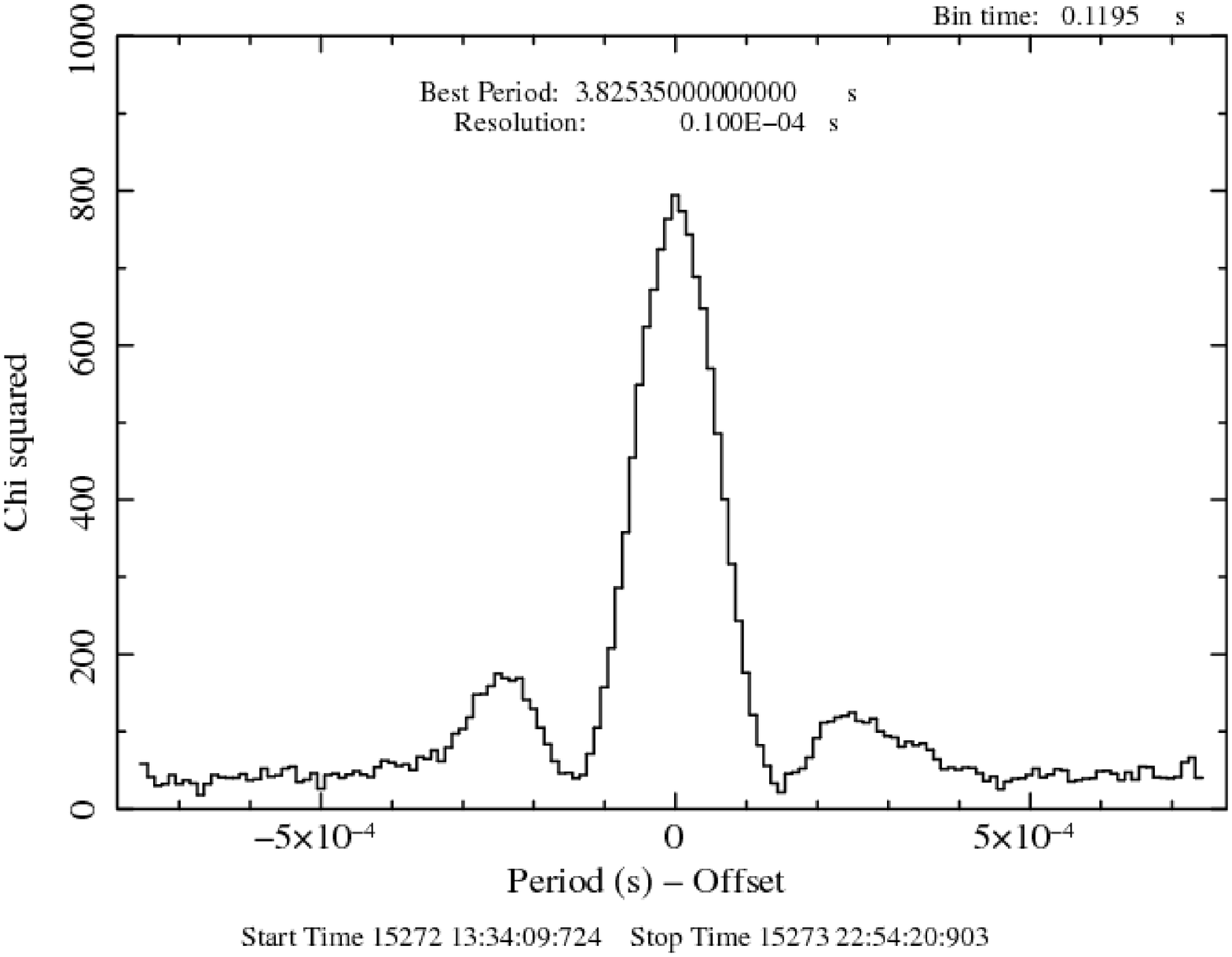}
\FigureFile(80mm,50mm){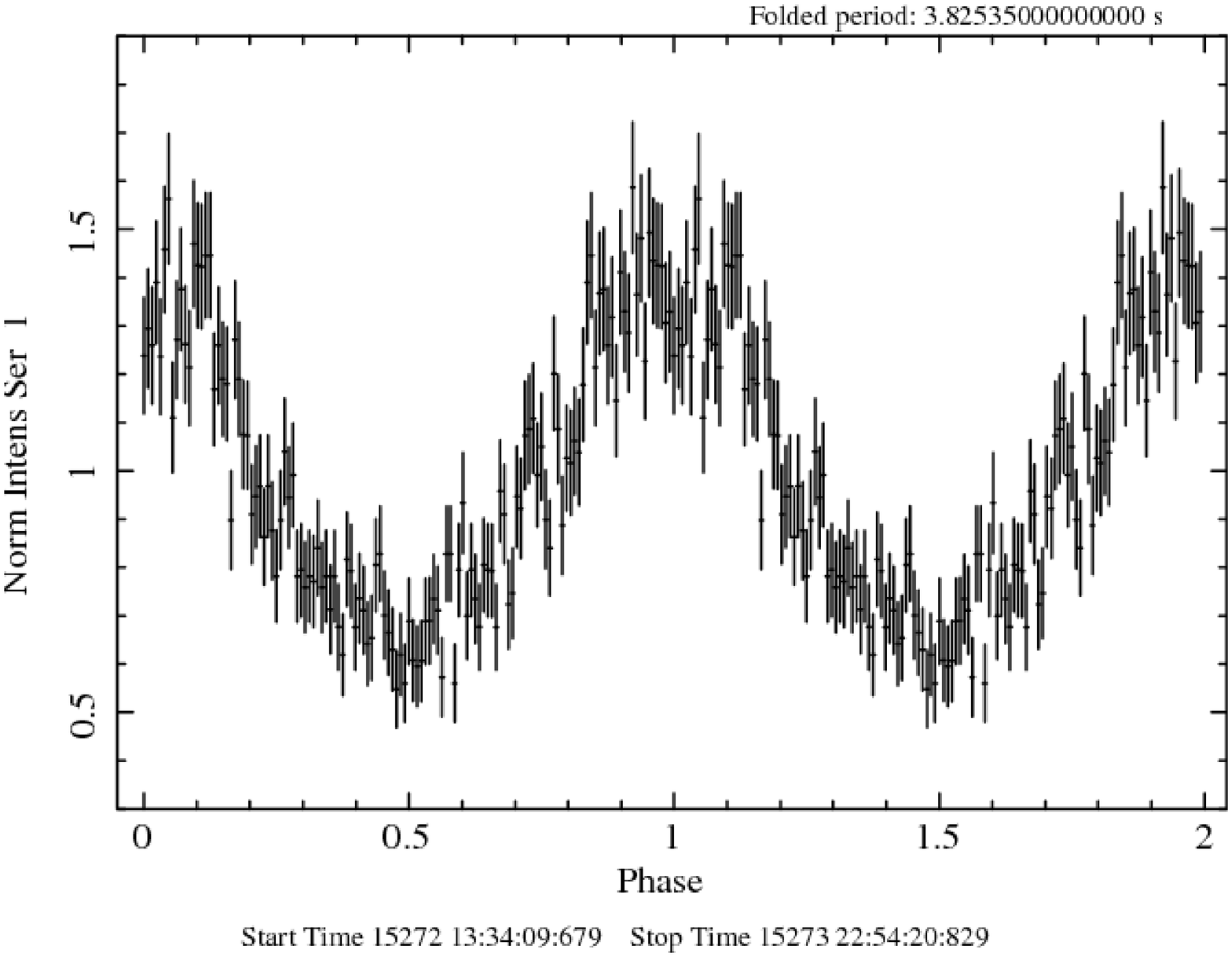}
\end{center}
 \caption{(a) Power spectrum of {\cxo} in the 1--10~keV band. A highly
 significant peak is detected at a frequency of
 0.2614~Hz. 
(b) Periodogram around the
 0.2614~Hz. The pulsation period is obtained to be
 3.82535$\pm$0.00005~s. (c) The light curve folded at this
 period. \label{fig:timing}}
\end{figure}

Since there is nearly no source flux below $\sim$1~keV, we have carried
out timing analysis in the band 1--10~keV. After barycentric correction
to the event file, we have created a light curve with the minimum time
resolution (73.4~ms), and have first made a power spectrum in the
frequency range below 0.5~Hz. The result is shown in
Fig.~\ref{fig:timing}(a).
A highly significant peak ($\sim$100~$\sigma$) appears at 0.2614~Hz. We
then have carried out epoch folding analysis near this frequency. The
resultant periodogram is shown in Fig.~\ref{fig:timing}(b). The
rotational period is obtained to be $P = 3.82535\pm 0.00005$~s.
No other period except for the harmonics
were significant.
Compared to the period obtained from the {\chan} cc mode data
\citep{2010ApJ...710..941H}, 3.82305$\pm$0.00002~s, the period becomes
longer by 0.00230$\pm$0.00005~s. The time has elapsed since the {\chan}
observation (beginning at 2009 Jan 25 06:55:08) by 416.264676~d, thereby
the average period derivative is obtained to be $\dot P = 6.40\pm
0.14\times 10^{-11}$ s~s$^{-1}$. Figure~\ref{fig:timing}(c) shows the
light curve folded at the best spin period. The pulse profile is similar
to that obtained by \citet{2010ApJ...710..941H} including the pulse
fraction.

\section{Discussion}

From the timing analysis of the {\xmm} pn data, we have obtained
$\dot P = 6.40\pm 0.14\times 10^{-11}$ s~s$^{-1}$. Together with $P =
3.82535$~s, we can derive the spin-down luminosity ($\dot E = 3.9\times
10^{46}\dot P P^{-3}$~erg~s$^{-1}$),
characteristic age ($t_c = P/(2\dot P)$~s), and
the dipole surface magnetic field ($B_s = 3.2\times 10^{19}\sqrt{P\dot
P}$~G) to be $4.5\times 10^{34}$~ergs~s$^{-1}$, $9.5\times 10^{2}$~yr, and
$5.0\times 10^{14}$~G, respectively.  All these estimated parameters are
within the range of the known AXPs, and we thus conclude that {\cxo} is
a new magnetar, together with the large photon index of its spectrum.
The ratio of 2--10~keV luminosity ($L_x$) to $\dot E$ is 0.4,
which is much smaller than the typical magnetars.
PSR~J1846$-$0258 is a radio pulsar with $B_s$
larger than the critical magnetic field
which shows the $L_X/\dot E$ of 0.2 \citep{2003ApJ...582..783H}
including its pulsar wind nebula (PWN),
and should be a key source
to connect magnetars and conventional radio pulsars.
{\cxo} thus closely resembles PSR~J1846$-$0258 
and a new key source between magnetars and radio pulsars.

Note that this source is the youngest AXP and the second youngest
magnetar so far, in the next place of SGR~1806$-$20 ($t_c = 0.22$ kyr;
\cite{2005ApJ...628..938M}).  Another important point is that this
magnetar is associated with a young SNR. \citet{2009PASJ...61S.197N}
obtained the ionization age of the thermal plasma associated with CTB37B
is $650^{+2500}_{-300}$~yr. Possible association of CTB37B with the
historical SNR SN~393 has long been discussed
\citep{1975Obs....95..190C,2002ISAA....5.....S}. The characteristic age
we obtained is consistent with these discussions.
\citet{2009ApJ...707L.148V} discovered a pulsar wind nebula around the
second youngest AXP, 1E~1547.0$-$5408 ($t c = 1.4$ kyr:
\cite{2007ApJ...666L..93C}). {\cxo} is now a good target to search for a
PWN, which will be a future work with better spatial
resolution. This will lead to better understanding of the young
magnetars.

Some magnetars show X-ray flares and long term variability, and hence we
have investigated whether {\cxo} has X-ray time variability.
The absorbed 2--10~keV flux is
$(1.1\pm0.2)\times 10^{-12}$~ergs~cm$^{-2}$s$^{-1}$ on
2007 Feb. 2 by {\chan} \citep{2009PASJ...61S.197N},
$1.8\times 10^{-12}$~ergs~cm$^{-2}$s$^{-1}$ on 2009 Jan. 25 by {\chan}
\citep{2010ApJ...710..941H},
and $(1.51\pm0.03)\times 10^{-12}$~ergs~cm$^{-2}$s$^{-1}$
on 2010 Mar. 17-18 by {\xmm} by this work.
We thus concluded that
{\cxo} showed significant time variability in these years.

\bigskip

We acknowledge Jacco Vink for his fruitful discussions.
This work was supported in part by
Grant-in-Aid for Scientific Research
of the Japanese Ministry of Education, Culture, Sports,
Science
and Technology, No.~22684012 (A.~B.).


\begin{thebibliography}{}
\bibitem[Aharonian et al.(2006)]{2006ApJ...636..777A}
 Aharonian, F., et al.\ 2006, \apj, 636, 777 
\bibitem[Aharonian et al.(2008)]{2008A&A...486..829A}
 Aharonian, F., et al.\ 2008, \aap, 486, 829 
\bibitem[Anders \& Grevesse(1989)]{1989GeCoA..53..197A}
Anders, E., \& Grevesse, N.\ 1989, \gca, 53, 197
\bibitem[Bamba et al.(2005)]{2005ApJ...621..793B}
Bamba, A., Yamazaki, R., 
Yoshida, T., Terasawa, T., \& Koyama, K.\ 2005, \apj, 621, 793
\bibitem[Camilo et al.(2007)]{2007ApJ...666L..93C} Camilo, F., Ransom, 
S.~M., Halpern, J.~P., \& Reynolds, J.\ 2007, \apjl, 666, L93 
\bibitem[Caswell et al.(1975)]{1975A&A....45..239C}
Caswell, J.~L., Murray, J.~D., Roger, R.~S., Cole, D.~J., \& Cooke,
D.~J.\ 1975, \aap, 45, 239
\bibitem[Clark \& Stephenson(1975)]{1975Obs....95..190C}
 Clark, D.~H., \& Stephenson, F.~R.\ 1975, The Observatory, 95, 190 
\bibitem[Duncan \& Thompson(1992)]{1992ApJ...392L...9D}
Duncan, R.~C., \& Thompson, C.\ 1992, \apjl, 392, L9
\bibitem[Enoto et al.(2010)]{enoto2010}
Enoto, T., Nakazawa, K., Makishima, K., Rea, N., Hurley, K.,
and Shibata, S.\ 2010, \apjl, submitted
\bibitem[Garmire et al.(2000)]{2000AJ....120.1426G}
Garmire, G., Feigelson, 
E.~D., Broos, P., Hillenbrand, L.~A., Pravdo, S.~H., Townsley, L., 
\& Tsuboi, Y.\ 2000, \aj, 120, 1426 
\bibitem[Halpern \& Gotthelf(2010)]{2010ApJ...710..941H}
 Halpern, J.~P., \& Gotthelf, E.~V.\ 2010, \apj, 710, 941 
\bibitem[Helfand et al.(2003)]{2003ApJ...582..783H}
Helfand, D.~J., 
Collins, B.~F., \& Gotthelf, E.~V.\ 2003, \apj, 582, 783
\bibitem[Jansen et al.(2001)]{2001A&A...365L...1J}
 Jansen, F., et al. \ 2001, \aap, 365, L1 
\bibitem[Kassim et al.(1991)]{1991ApJ...374..212K}
 Kassim, N.~E., Weiler, K.~W., \& Baum, S.~A.\ 1991, \apj, 374, 212 
\bibitem[Koyama et al.(2007)]{2007PASJ...59S..23K}
 Koyama, K., et al. \ 2007, \pasj, 59, 23 
\& Yonetoku, D.\ 2003, \apj, 585, 948 
\bibitem[Mereghetti et al.(2005)]{2005ApJ...628..938M}
 Mereghetti, S., et al. \ 2005, \apj, 628, 938 
\bibitem[Mitsuda et al.(2007)]{2007PASJ...59S...1M}
 Mitsuda, K., et al.\ 2007, \pasj, 59, 1 
\bibitem[Muno et al.(2007)]{2007MNRAS.378L..44M} Muno, M.~P., Gaensler, 
B.~M., Clark, J.~S., de Grijs, R., Pooley, D., Stevens, I.~R., 
\& Portegies Zwart, S.~F.\ 2007, \mnras, 378, L44 
\bibitem[Naik et al.(2008)]{2008PASJ...60..237N} Naik, S., et al.\ 2008, 
\pasj, 60, 237
\bibitem[Nakagawa et al.(2009)]{2009PASJ...61..109N} Nakagawa, Y.~E., 
Yoshida, A., Yamaoka, K., \& Shibazaki, N.\ 2009, \pasj, 61, 109 
\bibitem[Nakamura et al.(2009)]{2009PASJ...61S.197N}
 Nakamura, R., Bamba, A., Ishida, M., Nakajima, H., Yamazaki, R.,
 Terada, Y., P\"{u}hlhofer, G., \& Wagner, S.~J.\ 2009, \pasj, 61, 197 
\bibitem[Stephenson \& Green(2002)]{2002ISAA....5.....S}
 Stephenson, F.~R., \& Green, D.~A.\ 2002, Historical supernovae and their remnants, by F.~Richard Stephenson and David A.~Green.~International series in astronomy and astrophysics, vol.~5.~Oxford: Clarendon Press, 2002, ISBN 0198507666, 5,\bibitem[Str\"{u}der et al.(2001)]{2001A&A...365L..18S}
 Str\"{u}der, L., et al. \ 2001, \aap, 365, L18 
\bibitem[Thompson \& Duncan(1995)]{1995MNRAS.275..255T}
Thompson, C., \& Duncan, R.~C.\ 1995, \mnras, 275, 255
\bibitem[Thompson \& Duncan(1996)]{1996ApJ...473..322T}
Thompson, C., \& Duncan, R.~C.\ 1996, \apj, 473, 322 
\bibitem[Turner et al.(2001)]{2001A&A...365L..27T} Turner, M.~J.~L., et al.\ 2001, \aap, 365, L27 
\bibitem[Vink \& Bamba(2009)]{2009ApJ...707L.148V}
Vink, J., \& Bamba, A.\ 2009, \apjl, 707, L148

\end{thebibliography}
\end{document}